\documentclass{easychair}

\usepackage{amsmath}
\usepackage{amssymb}
\usepackage{hyperref}
\usepackage{graphicx}
\usepackage{booktabs}
\usepackage{makecell}
\usepackage[noend]{algpseudocode}
\usepackage{algorithm}
\usepackage{algorithmicx}
\usepackage{subcaption}
\usepackage{paralist}
\usepackage{float}
\usepackage{xcolor}

\newcommand*{\para}[1]{\noindent\textbf{#1}}
\renewcommand*{\P}{\mathcal{P}}

\begin{document}

\title{\textbf{Stateful Premise Selection\\
    by Recurrent Neural Networks}}

\titlerunning{Stateful Premise Selection with RNNs}

\author{
Bartosz Piotrowski\inst{1}\inst{2}\thanks{
	Supported by the \textit{AI4REASON} ERC Consolidator grant nr. 649043 and
	by the grant 2018/29/N/ST6/02903 of National Science Center, Poland.
	}
\and
Josef Urban\inst{1}\thanks{
	Supported by the \textit{AI4REASON} ERC Consolidator grant nr. 649043
	and by the Czech project AI\&Reasoning CZ.02.1.01/0.0/0.0/15\_003/0000466
	and the European Regional Development Fund.
	}
}

\authorrunning{Piotrowski and Urban}

\institute{
Czech Institute of Informatics, Robotics and Cybernetics, Prague, Czech Republic
\and
Faculty of Mathematics, Informatics and Mechanics, University of Warsaw, Poland
}

\maketitle

\begin{abstract}

In this work, we develop a new learning-based method for selecting facts
(premises) when proving new goals over large formal libraries. Unlike previous
methods that choose sets of facts independently of each other by their rank,
the new method uses the notion of \emph{state} that is updated each time a
choice of a fact is made. Our stateful architecture is based on recurrent
neural networks which have been recently very successful in stateful tasks such
as language translation. The new method is combined with data augmentation
techniques, evaluated in several ways on a standard large-theory benchmark, and
compared to state-of-the-art premise approach based on gradient boosted trees.
It is shown to perform significantly better and to solve many new problems.

\end{abstract}

\section{Introduction: Premise Selection over Large Libraries}
\label{intro}

\emph{Premise selection}~\cite{abs-1108-3446} is a critical task in automated
theorem proving (ATP) over large theories where typically only a small fraction
of the available facts is relevant for proving a new conjecture. One of the
main applications is in ITP/ATP \emph{hammers}~\cite{hammers4qed} that assist
ITP users to automatically discharge proof obligations in large formalizations.
Several heuristic approaches to premise selection, such as SiNE~\cite{HoderV11}
and MePo~\cite{MengP09}, as well as learning-based methods using hand-designed
features, have been developed so far. The latter include naive
Bayes~\cite{Urban06}, kernel methods~\cite{TsivtsivadzeUGH11}, k-nearest
neighbors \cite{Kaliszyk14,Kaliszyk15b} and gradient boosted trees
\cite{Piotrowski18}. This has been followed by neural architectures that learn
the features on their own \cite{Irving16,DBLP:journals/corr/abs-1807-10268}.

The learning-based premise selection methods have been so far based on the same
paradigm of \emph{ranking} the available facts (premises) \emph{independently}
with respect to the conjecture that is being proved. The highest-ranked facts
are then used together as axioms and given to the ATP systems together with the
conjecture. This approach, although useful and reasonably successful, does not
take into account an important aspect of the premise selection problem:
premises are \emph{not} independent of each other. There are important logical
relations among them. Some premises complement each other better when proving a
particular conjecture, while some highly-ranked premises might be just minor
variants of one another.

In this work, we first (Section~\ref{sec:nmt}) propose the recurrent neural
network (RNN) encoder-decoder model~\cite{cho2014learning} as a suitable
stateful approach for premise selection and we describe the RNN architecture we
have chosen for this task. In Section~\ref{sec:data}, we develop several data
augmentation methods that help training the RNNs for the premise selection
task. Section~\ref{sec:eval} describes the experimental evaluation, and
Section~\ref{sec:conclusion} discusses the results.

\section{Premise Selection and Neural Machine Translation}
\label{sec:nmt}

Over the last few years, powerful methods for learning sequences of conditional
stateful decisions have been developed in machine translation of natural
languages. In \textit{neural machine translation} (NMT)~\cite{cho2014learning}
the source sentence is encoded as hidden vector representation by the
\emph{encoder}, and the translated target sentence is produced word-by-word by
the \emph{decoder}.  Each translated word is conditioned not only on the source
sentence but also on the \textit{previously decoded words}. Words and phrases
in natural languages are related in many ways, and such relations have to be
taken into account for the produced sequence of words to be a sensible,
grammatically correct sentence.  Successful NMT methods using recurrent
encoder-decoder architectures~\cite{NIPS2014_5346} are explicitly based on a
notion of a learned \emph{hidden state} that is updated with each produced
word. This corresponds to our requirement of \emph{stateful premise selection}:
We want to have a (learned) hidden state after selecting a particular fact with
particular mathematical content which should not be repeated in the following
facts but rather suitably complemented by them to justify the conjecture.

Another aspect of neural machine translation that is relevant in premise
selection is \textit{the multiplicity of correct outputs}. In translation,
there are often multiple correct translations of a given sentence that deliver
its \emph{meaning} (perhaps more or less clumsily). Similarly, in mathematics,
there is typically no single, golden proof of a conjecture. Often there are
many different proofs that use various sets of premises and various sequences
of inferences. NMT methods accommodate the multiplicity of possible outputs --
typically by using \textit{beam search} \cite{Freitag17}. Such mechanisms seem
directly usable also in premise selection and proof search.  NMT systems have
already been successfully applied in autoformalization and symbolic settings
\cite{DBLP:conf/mkm/WangKU18, DBLP:journals/corr/abs-1911-04873, Lample19}.


\vspace{1mm}
\para{Our Recurrent Neural Architecture:}
There are various state-of-the-art neural sequence-to-sequence architectures
that can be applied to modelling premise selection. Although ultimately a
custom architecture could be designed to capture all aspects of this task, our
initial choice was to experiment with an existing established implementation of
a neural machine translation system. We have chosen the OpenNMT toolkit
\cite{opennmt17}. It implements the LSTM \cite{Hochreiter97} recurrent cells
and several state-of-the-art techniques, including the attention mechanisms
\cite{Luong15} and beam search \cite{Freitag17}. It has proven to be very good
in natural language translation and related tasks \cite{opennmt17}.

We have decided to use the default parameters for training OpenNMT on our tasks
-- in this work, we mainly investigate the influence of various forms of
training data (Section \ref{sec:data}) on the predictive performance. The more
important values of the OpenNMT hyperparameters chosen by us are as follows:
the number of training epochs: 100000, the size of encoder's and decoder's LSTM
cells: 2 layers of 500 units, word embedding size: 500.  Additionally, we have
used the attention mechanism by Luong \cite{Luong15}.

\section{Data, Their Representation and Augmentation}
\label{sec:data}

A recurrent NMT system is trained on pairs of \textit{source} and
\textit{target} sequences. In our case, the source is a statement of a theorem
and the target is a list of names of its premises. There are multiple ways how
to transform ATP proofs to training examples of such form, and it is not clear
which way is the best for training the RNN. In this section, we describe
several methods of constructing the training examples\footnote{All the data
used in this work along with scripts allowing reproduction of the experiments
are available at:
\url{https://github.com/BartoszPiotrowski/stateful-premise-selection-with-RNNs}}.
This includes the following topics: (i) representation of the conjecture as a
sequence (Section \ref{sec:source_format}), (ii) ordering of the premises into
a sequence (Section \ref{sec:target_order}), (iii) using subproof data for
augmenting the training data (Section \ref{sec:augmentation_with_subproofs}),
and (iv) oversampling of rare premises (Section \ref{sec:oversampling}).

\subsection{Initial Data for Training RNNs}

The experimental data originate from the Mizar Mathematical Library
(MML) \cite{Grabowski10} translated~\cite{Urban06} to the TPTP
language~\cite{Sutcliffe10}. 
We use the MPTP2078 benchmark~\cite{abs-1108-3446} -- a subset of 2078 Mizar
theorems. Using the ATPboost \cite{Piotrowski18} system we have initially
proved as many of the MPTP2078 problems as possible, recording each distinct
proof. ATPboost in turn relies on the E prover \cite{Schulz13} and the XGBoost
machine learning system \cite{xgboost} using gradient boosted trees for premise
selection. 24087 different proofs of 1469 theorems were found in total. The
number of different proofs per theorem ranged between 1 and 265 (on average
$16.4$). The proofs used in total 2227 different premises. Each proof used
between 1 and 50 of the premises (on average $11.5$).
Each proof determines a pair $(t,\{p_1,p_2,\ldots,p_n\})$, where the first
element $t$ is the proved theorem and the second element is the set of premises
$p_i$ used in the proof. These pairs constitute examples for training a machine
learning model to propose useful premises for theorems.  The 1469 theorems that
have an ATP proof were randomly split in proportions 0.75 and 0.25 into
training and testing parts. The 1100 training theorems with their proofs
resulted in 18361 training pairs. From the set of remaining 369 theorems we
filtered out those which contained in all their proofs premises not appearing
in the training set. This yields our testing set of 310 theorems.

\subsection{Representation of the Statements}
\label{sec:source_format}

The simplest way of constructing the source sequences of the examples is just
using tokenized statements in standard TPTP syntax. We label this type of
source as \texttt{standard}.  The tokenized TPTP statements can also be
transformed into other formats. A suitable one is the Polish prefix notation
(labeled as \texttt{prefix}), as shown in Table \ref{tab:prefix_notation}. The
motivation is that this format is more compact as the formulas do not contain
brackets and commas. In our case, the average length of a \texttt{standard}
source training sequence is 81, whereas for \texttt{prefix}, it is only 39.
This may be useful for NMT architectures that suffer from long input (and
output) sequences \cite{Cho14}. Related work using NMT in symbolic setting
reports improvements when using prefix notation \cite{DBLP:conf/mkm/WangKU18}.

\begin{table}
\centering
\caption{An example of a Mizar statement translated to TPTP,
	tokenized (\texttt{standard}), and additionally transformed to prefix
	notation (\texttt{prefix}).}
\label{tab:prefix_notation}
\centering
\vspace{-3mm}
\begin{tabular}{l@{\hskip 4mm}l}
\toprule
Mizar:             & \texttt{ for A being set st A is empty holds A is finite} \\
TPTP:              & \texttt{ ![A] : (v1\_xboole\_0(A) => v1\_finset\_1(A))} \\
\texttt{standard}: & \texttt{ ! [ A ] : ( v1\_xboole\_0 ( A ) => v1\_finset\_1 ( A ) )} \\
\texttt{prefix}:   & \texttt{ ! A => v1\_xboole\_0 A v1\_finset\_1 A } \\
\bottomrule
\end{tabular}
\end{table}

\subsection{Ordering  of the Premises}
\label{sec:target_order}

In the abstract premise selection task the order of the premises is not
constrained in any way.  In practice, ATP systems may be influenced by order of
the premises given in the input. More importantly, existing efficient learning
methods that are capable of capturing dependencies among the elements of
sequences (such as RNNs, used this work) are sensitive to the order of the
elements (premises in our case).
Preferably, we want to train on examples that illustrate dependencies between
the premises. On the other hand, we do not want  to rely too much on a
particular order of premises in the target sequence. We propose several
possible approaches.

\vspace{1mm}
\para{Permutations:}
As a baseline approach, we permute the target premises randomly, thus not
passing to the recurrent neural model any additional information about the
order of the premises. We either produce only one permutation
(\texttt{permuted}) or 100 of them (\texttt{permuted\_100}).

\begin{figure}
	\centering
	\includegraphics[width=\textwidth]{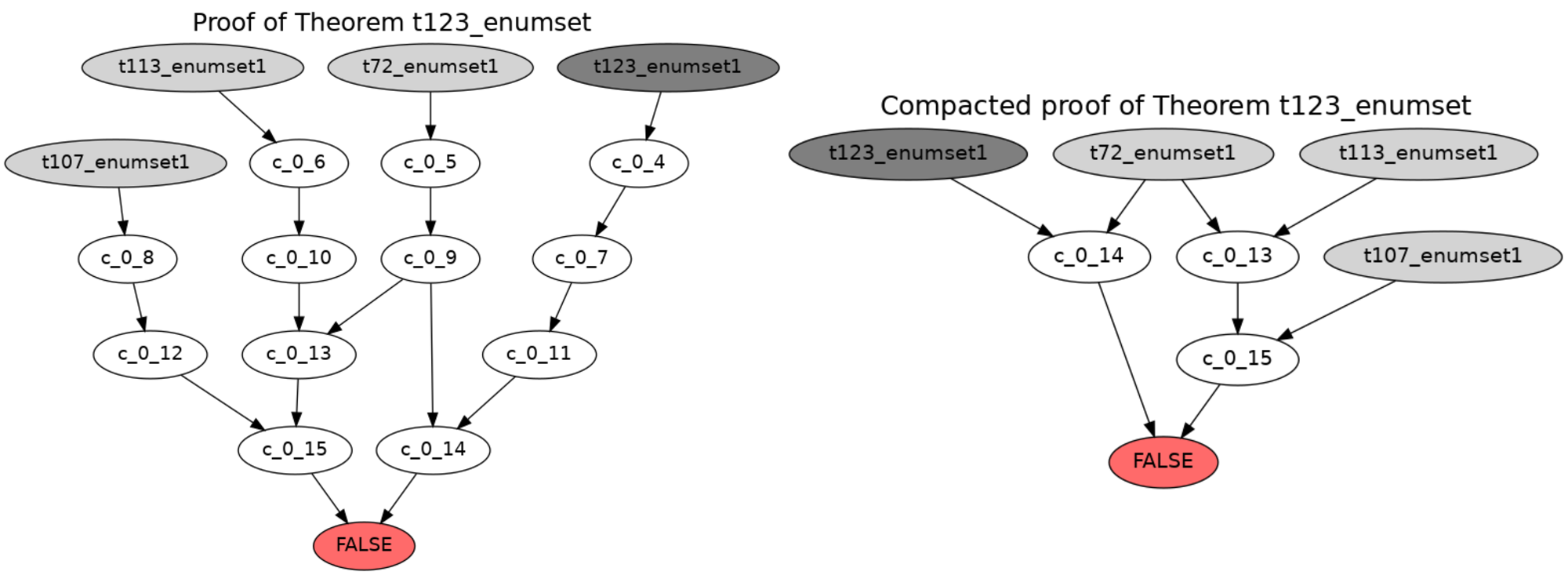}
	\caption{\label{fig:tree}
	Trees representing a refutational proof of Mizar theorem
	\texttt{t123\_enumset} (left) and its compacted version (right). Light-grey
	nodes are the input premises and a dark-grey node represents the (negated)
	conjecture. Nodes prefixed by \texttt{c\_0} are intermediate lemmas.
      }
\vspace{-4mm}
\end{figure}

\vspace{2mm}
\para{Permutations Preserving the Proof Tree:}
Each proof produced by a refutational 
prover (such as E) is a tree (more precisely, a DAG), with \texttt{FALSE} in
its root and the premises and the conjecture in its leaves. See Figure
\ref{fig:tree} for an example.
We produce a compacted version of the proof by removing all intermediate nodes
that have only one parent\footnote{Parent/child terminology is used in such a
way that children are derived from its parents.} (right side of Figure
\ref{fig:tree}).  The premises \texttt{t72\_enumset1} and
\texttt{t113\_enumset1} there interact directly, resulting in an intermediate
lemma \texttt{c\_0\_13}. This lemma subsequently interacts with
\texttt{t107\_enumset1}.  Putting \texttt{t72\_enumset1} and
\texttt{t113\_enumset1} closer together than \texttt{t107\_enumset1} and
\texttt{t72\_enumset1} tells the learner how much the premises interact with
each other. This idea is implemented in the following way. Each tree induces
its \emph{nested list representation}, which we define recursively in the
following way:
(i) list representation of a tree rooted in a non-leaf node $n$ is a list of
list representations of parents of $n$,
(ii) list representation of a leaf node is its label.
For instance, \texttt{[[t123, t72], [[t72, t113], t107]]} is a list
representation of the tree from Figure \ref{fig:tree}.  Each tree has many list
representations depending on how the elements of the lists are ordered.

We say that a sequence $s$ of labels of the leaves \emph{respects the tree} if
$s$ resulted from the flattening of a list representation of the tree.
(In the example shown in Figure \ref{fig:tree} the sequence
(\texttt{t123}, \texttt{t72}, \texttt{t72}, \texttt{t113}, \texttt{t107})
respects the tree, but the sequence
(\texttt{t123}, \texttt{t107}, \texttt{t72}, \texttt{t113}, \texttt{t72})
does not.)
%
Such sequences have the property of keeping closer the premises that interacted
closer. Note that the sequences may contain repetitions, as in the example
above, and each tree has many sequences respecting it.  For each proof tree, we
remove from its sequences the conjecture.  We take either only one such
sequence for each proof or (up to) 100 different sequences, which yields the
\texttt{permuted\_tree} and \texttt{permuted\_tree\_100} sets of training
examples.

\vspace{1mm}
\para{ATP Induced Order:}
We also experiment with using the proofs as linearized by E prover.
For a given E proof $\P$ we first order its internal lemmas: \texttt{lemma\_1}
$<_L$ \texttt{lemma\_2} iff \texttt{lemma\_1} appears in $\P$ (linearized by E)
before \texttt{lemma\_2}.
Then we define a non-strict ordering of the premises of $\P$: $p_1 \leq_P p_2$
iff the $<_L$-minimal child of $p_1$ is smaller than the $<_L$-minimal child of
$p_2$, where both children are taken from the \emph{compacted} tree of $\P$.
For our example proof of \texttt{t123\_enumset} we have:
\texttt{c\_0\_13} $<_L$ \texttt{c\_0\_14} $<_L$ \texttt{c\_0\_15}.
Hence \texttt{t72} $\leq_P$ \texttt{t107}
(since $\text{min}_L(\{\texttt{c\_0\_14}, \texttt{c\_0\_13}\}) <_L \texttt{c\_0\_15}$),
and \texttt{t113} $=_P$ \texttt{t107}
(since $\text{min}_L(\{\texttt{c\_0\_14}, \texttt{c\_0\_13}\}) =
\texttt{c\_0\_13}$). We break the ties randomly.
Different E proofs of one theorem may result in different premise orderings.
This way of ordering premises in the target of the examples is labeled as
\texttt{order\_from\_proof}.


\subsection{Augmentation with Subproof Data}
\label{sec:augmentation_with_subproofs}

We can also augment the training data by extracting many subproofs from the
original training proofs.
For this, we use the intermediate lemmas from the compacted representations of
proofs that are not derived from the negated conjecture.
The pairs $(l, \{p_1,p_2,\ldots,p_n\})$
where $p_i$ are all premise ancestors of lemma $l$
constitute additional examples that can be used for augmenting the main
training data. From the subproofs of the training theorems we extracted 46094
such different training pairs. The data set that includes these examples
together with the main ones is marked as \texttt{augmented}.
The experiments with sublemmas only are described in Appendix
\ref{sec:subproofs}.

\subsection{Oversampling Rare Examples}
\label{sec:oversampling}

Some premises appear frequently in the training examples, while some are rare.
\emph{Oversampling} is a general method that often improves the performance of
neural architectures on imbalanced data~\cite{buda2018systematic}.  We
experiment with a simple oversampling scheme: training examples that contain
rare premises are used multiple times.  More precisely, for an example $e = (t,
P) \in \mathcal{T}$, where $\mathcal{T}$ is the training set, we define the
\emph{occurrence rate} (OR) of $e$ as:
\vspace{-3mm}
\[
	\text{OR}(e)
	= \frac{1}{|P|}\sum_{p \in P} \text{OR}_\text{premise}(p),
	\hspace{0.2cm} \text{where} \hspace{0.2cm}
	\text{OR}_\text{premise}(p) = \frac{|\{(t, P) \in \mathcal{T} \colon
	p \in P \}|}{\sum_{(t, P) \in
	\mathcal{T}} |P|}.
\vspace{-2mm}
\]
The idea is simple: OR of an example is the average OR$_\text{premise}$ of its
target premises ($P$). OR$_\text{premise}$ measures how often a premise appears in
all the targets of all the training examples.

The set of training examples $\mathcal{T}$ is split into $10$ evenly sized
chunks $\mathcal{T}_1, \mathcal{T}_2, \ldots, \mathcal{T}_{10}$ according to
their occurrence rate so that a higher index of $\mathcal{T}_i$ implies lower OR:
\vspace{-2mm}
\[
	x \in \mathcal{T}_i \wedge y \in \mathcal{T}_j \wedge i < j
	\Rightarrow
	\text{OR}(x) < \text{OR}(y).
\vspace{-2mm}
\]
Each example $e \in \mathcal{T}_i$ is oversampled $i$ times: the more rare
premises an example $e$ contains the more often $e$ appears in the oversampled
training set. This scheme was applied both to the main training set and to the
augmented one (described in Section \ref{sec:augmentation_with_subproofs}),
resulting in data sets \texttt{oversampled} and
\texttt{augmented\_oversampled}.

\section{Experimental Evaluation}
\label{sec:eval}

We train\footnote{We train on a single GeForce RTX 2080 Ti GPU. Each training
took between 2 and 4 hours.}  and evaluate RNNs using the OpenNMT toolkit with
its default hyper-parameters (Section \ref{sec:nmt}) on the various
premise-selection data described in Section \ref{sec:data}. When evaluating on
the testing sets, we use OpenNMT's beam search with width 10 to get for each
conjecture its 10 most probable sequences of premises.  We want to compare the
results also with state-of-the-art premise selection based on gradient boosted
trees using the XGBoost toolkit.  For that, we use the features and settings
developed in our ATPboost system~\cite{Piotrowski18}.  As explained in
Section~\ref{intro}, the training data used for XGBoost are unordered. XGBoost
produces a ranking of the premises and we use several segments of the
top-ranked premises for the XGBoost evaluation.  While OpenNMT needs only
positive examples, XGBoost also needs negative examples. We produce them by
sampling negatives randomly, which performed well in ATPboost.

To allow a meaningful comparison of the two approaches, we shorten the rankings
produced by XGBoost  according to the lengths of the sequences produced by
OpenNMT for a given conjecture. In more detail: if $\widehat{R}$ is a ranking
produced by XGBoost and
$\widehat{P}_1, \widehat{P}_2, \ldots, \widehat{P}_{10}$
are sequences of premises produced by OpenNMT, we take
$\widehat{R}_1, \widehat{R}_2, \ldots, \widehat{R}_{10}$
to be the top slices of the ranking $\widehat{R}$ of lengths
$|\widehat{P}_1|, |\widehat{P}_2|, \ldots, |\widehat{P}_{10}|$,
respectively. These 10 top slices are treated as predictions from the XGBoost
system and compared with the OpenNMT predictions.
We always do both the standard \emph{ML evaluation} and the \emph{ATP evaluation}.

\vspace{2mm}
\para{ML Evaluation:}
The {Jaccard index} and {Coverage} metrics are used, defined as below:
\vspace{-2mm}
\[
	\textnormal{Jaccard}(A, B) = \frac{| A \cap B |}{|A \cup B|},
	\hspace{1cm}
	\textnormal{Coverage}(A, B) = \frac{| A \cap B |}{|B|}.
\vspace{-2mm}
\]
Each theorem $T$ from the test set is associated with $n_T$ sets of premises
$P^T_1, P^T_2, \ldots, P^T_{n_T}$ which were used as axioms in its $n_T$ known
proofs and with 10 sets of premises $\widehat{P}^T_1, \widehat{P}^T_2, \ldots,
\widehat{P}^T_{10}$ predicted by a given machine learning model.  We estimate
the quality of the predictions with $\text{Jaccard}(\bigcup_{i}
\widehat{P}^T_i, \bigcup_{i} P^T_i)$ and $\text{Coverage}(\bigcup_{i}
\widehat{P}^T_i, \bigcup_{i} P^T_i)$.  The Jaccard metric emphasises the
\textit{precision} of the prediction: it measures how much the predicted
premises intersect with the premises used in the known proofs. At the same time
the score decreases when the predicted set is large. The Coverage does not
penalize large predicted sets. It takes into account the fact that \textit{true
positives} are typically more important than \textit{true negatives}. The
prover may deal with some redundant axioms while the lack of relevant premises
may make conjectures unprovable.

\vspace{2mm}
\para{ATP Evaluation:}
\label{sec:atp_eval}
The simple ML metrics may not directly translate to ATP performance. They
compare unions of premises, whereas an ATP is run for each premise selection
separately.  Additionally, during the ATP evaluation new proofs are often
discovered.  Such new proofs are not taken into account by the similarity
metrics.

We perform an ATP evaluation using the E prover~\cite{Schulz13} run with a time
limit of 10 s and a memory limit of 2 GB, keeping the rest of the settings in
its default values. This limits the power of the prover, preventing e.g., its
own axiom pruning methods such as SInE~\cite{HoderV11}.
To establish a simple ATP baseline, E prover was run for all the testing
theorems with all available premises as axioms, proving 9\% of the theorems.
For a given machine learning method and for each testing theorem $T$ we run 11
proof attempts. One for each of the 10 sets of predicted premises
$\widehat{P}^T_1, \widehat{P}^T_2, \ldots, \widehat{P}^T_{10}$ and one with
$\bigcup_{i} \widehat{P}^T_i$.

\section{Results and Discussion}
\label{sec:conclusion}

\subsection{Source and Target Combinations}
\label{sec:srctgt}
First, we evaluate combinations of the statement format (\texttt{standard} or
\texttt{prefix} -- Section \ref{sec:source_format}) and orderings of the target
premises (Section \ref{sec:target_order}). The results are shown in
Table~\ref{tab:results_source_targets}.
\begin{table}
\caption{\label{tab:results_source_targets}
Performance of the neural model on the test set, trained on examples with
different formats of source sequences (statements) and differently ordered
target sequences (premises), expressed with the similarity metrics ({Jaccard
index} and {Coverage}) and with ATP success rate.}
\centering
\vspace{-2mm}
\begin{tabular}{lccc@{\hspace{0.3cm}}ccc}
\toprule
&\multicolumn{6}{c}{\textbf{Source format}} \\
\cmidrule[0.05em](r){2-7}
& \multicolumn{3}{c}{\texttt{standard}} & \multicolumn{3}{c}{\texttt{prefix}} \\
\cmidrule[0.05em](r){2-4} \cmidrule[0.05em](r){5-7}
	\textbf{Target ordering} &
	{{Jaccard}} & {{Coverage}} & {{Proved}} &
	{{Jaccard}} & {{Coverage}} & {{Proved}} \\
\cmidrule[0.05em](r){1-1}
\cmidrule[0.05em](r){2-4}
\cmidrule[0.05em](r){5-7}
\texttt{permuted} 				& 0.18 & 0.39 & 0.25 & 0.20 & 0.36 & 0.23 \\
\texttt{permuted\_100} 			& 0.09 & 0.14 & 0.10 & 0.18 & 0.26 & 0.19 \\
\texttt{permuted\_tree} 		& 0.16 & 0.23 & 0.16 & 0.17 & 0.25 & 0.18 \\
\texttt{permuted\_tree\_100} 	& 0.04 & 0.05 & 0.03 & 0.11 & 0.15 & 0.09 \\ 
\texttt{order\_from\_proof} 	& 0.22 & 0.46 & 0.29 & 0.22 & 0.43 & 0.27 \\
\bottomrule
\end{tabular}
\end{table}
The first, simplest way of ordering premises -- \texttt{permuted} -- performs
well: in combination with the two formats of source statements it resulted in
predictions with ATP success rates 25\% and 23\%.  Many permutations of the
same target sequence are bad for the NMT learner: \texttt{permuted\_100}
performed much worse than \texttt{permuted}. Using multiple permutations of the
target sequences was motivated by the fact that the order of premises should
not matter in the abstract premise selection task.  The recurrent neural
network, however, likely sees them as \emph{contradictory data} detrimental to
its training.

The \texttt{permuted\_tree} ordering
was meant to reflect the distance of interaction between the premises.
It however 
performed much worse than \texttt{permuted}. This may be caused by many
repetitions in the target sequences, which also increase their lengths, making
the task for the neural decoder more difficult. These repetitions appear
because of the repeated premises in the leaves of the proof trees.  Similarly
to  \texttt{permuted\_100}, adding multiple permutations
(\texttt{permuted\_tree\_100}) is detrimental also in the case of
\texttt{permuted\_tree}.

The best performing way of ordering the premises for the NMT learner is to use
the \texttt{order\_from\_proof}.  This is true both for the similarity metrics
and ATP performance and in combination with both formats of the conjecture.
This likely means that extracting the premise ordering from the proofs brings
useful and consistent information which the neural model is able to take
advantage of during the training.

There is only a small difference between the ATP performance of the
\texttt{standard} and \texttt{prefix} encoding of the conjecture.
This is somewhat surprising in the context of related work
\cite{DBLP:conf/mkm/WangKU18}, where prefix notation is useful for NMT
architectures.  Shorter sequences should be easier to process by the recurrent
encoder.  In our case, however, the different structure of the \texttt{prefix}
statements seems to reduce the benefit of conciseness.

\subsection{Augmentation with Subproof Data and Oversampling}

Next, we use the best performing combination (from now on called
\texttt{basic}) from Section~\ref{sec:srctgt} for evaluating the augmentation
and oversampling methods (Sections \ref{sec:augmentation_with_subproofs} and
\ref{sec:oversampling}).  The results are shown in Table
\ref{tab:results_aug_over_tptp}, which also contains the XGBoost results.

\begin{table}
\caption{\label{tab:results_aug_over_tptp}
Performance of the NMT and XGBoost systems trained on examples augmented with
subproof data and with oversampling applied.  Non-modified data set is denoted
as \texttt{basic}.  The examples used \texttt{standard} source format and
\texttt{order\_from\_proof} ordering of the target sequences.
}
\centering
\vspace{-2mm}
\begin{tabular}{l@{\hspace{0.2cm}}ccc@{\hspace{0.3cm}}ccc@{\hspace{0.2cm}}}
\toprule
	&\multicolumn{6}{c}{\textbf{Machine learning system}} \\
\cmidrule[0.05em](r){2-7}
&\multicolumn{3}{c}{\texttt{NMT}} & \multicolumn{3}{c}{\texttt{XGBoost}} \\
	\cmidrule[0.05em](r){2-4} \cmidrule[0.05em](r){5-7}
\textbf{Training data} & Jaccard & Coverage & Proved & Jaccard & Coverage & Proved \\
\cmidrule[0.05em](r){1-1}
\cmidrule[0.05em](r){2-4}
\cmidrule[0.05em](r){5-7}
\texttt{basic}                    & 0.22 & 0.46 & 0.29 & 0.26 & 0.56 & 0.25 \\
\texttt{oversampled} 	          & 0.21 & 0.48 & 0.31 & 0.24 & 0.61 & 0.30 \\ 
\texttt{augmented} 		          & 0.27 & 0.51 & 0.40 & 0.26 & 0.51 & 0.27 \\
\texttt{augmented\_oversampled}   & 0.26 & 0.47 & 0.39 & 0.25 & 0.51 & 0.31  \\ 
\bottomrule
\end{tabular}
\end{table}
Oversampling trained the learner using data with changed distribution -- less
frequent premises were appearing more often. This made the predictions more
diverse and less precise compared to \texttt{basic}. This is reflected in the
change of the similarity metrics: Jaccard index decreased and Coverage
increased. Importantly, oversampling translates to better ATP performance of
the NMT predictions.

Augmentation with subproof data improved the ATP performance by a large margin
of 11\% points over \texttt{basic}. It means that the RNN is helped a lot by
the additional subproof training data
despite their slightly different origin and shape (internal clauses instead of
input formulas).
The last row of Table~\ref{tab:results_aug_over_tptp} shows the result of
applying oversampling on top of the augmented training set.  This does not
improve the NMT performance compared to \texttt{augmented}.

ATP performance of XGBoost was worse than that of NMT for all 4 data sets, and
the best XGBoost ATP result (0.31) is significantly (29\%) worse than the best
NMT ATP result (0.40).
However, XGBoost tends to show better values than NMT in similarity metrics.
This can be explained by the following effects:
\begin{compactenum}
\item The initial data come from ATPboost -- a meta-system using XGBoost. The
	XGBoost predictions in our experiments may be correlated with the initial
	testing set.
\item Even though XGBoost achieves a higher similarity between unions
  of the predicted premises and the premises used in the known proofs, the
  recurrent neural network wins with its diverse (but stateful and therefore
  complementary) predictions for a given conjecture.  I.e., when making several
  ATP attempts, it seems better to use several plausible premise sets (proof
  ideas) that are orthogonal to one another, rather than making incremental
  additions to the initial set of premises. This is the effect that we wanted
  to achieve with RNN-based premise selection; indeed, it makes a significant
  difference. XGBoost instead just extends its single most plausible set of
  premises more and more according to the single ranking of premises. It seems
  nontrivial to instruct XGBoost to produce multiple alternative rankings with
  good complementary properties in the same way as RNN does.
\end{compactenum}

The NMT predictions are quite orthogonal to those from XGBoost. In all the
experiments related to the results shown in Table
\ref{tab:results_aug_over_tptp} there were 167 theorems proved with predictions
from NMT and 142 theorems proved with use of XGBoost. The size of the
intersection of these sets is 121, and there were 46 theorems which proofs were
found with NMT but not with XGBoost.

In Appendix \ref{sec:examples} we show some examples of predictions from NMT and XGBoost.



\section{Conclusions and Future Work}

We have shown that state-of-the-art recurrent neural architectures -- designed
originally for natural language tasks such as machine translation -- are very
useful for premise selection.  In particular, they can be used to implement (i)
stateful / conditional premise selection and (ii) beam search with multiple
output sequences that may differ a lot while being meaningful as a whole.  Our
experiments show significant improvement over the state of the art obtained by
such methods.  We have also developed several data representation and
augmentation methods that result in additional improved performance of both the
old and new premise selection methods.

NMT architectures are also more natural in some aspects.
There is no need for hand-designed features of formulas and no need to
construct negative training examples.  This is important because in
theorem proving it is often difficult (or impossible) to say that a
particular selection of premises \emph{cannot} lead to a proof.
Once the recurrent neural network is
trained, it directly outputs the most probable sequences of candidate premises -- not
just their rankings.
We have used 10 most probable sequences for the experiments described
here, but larger numbers can be used and given to ATPs, depending on available
resources.

Future work in this direction includes, for example, tighter integration
between the ATPs and the neural network, so that the prover can take advantage
of the order in which the premises are presented. Neural network research is
advancing quickly and experiments with other stateful neural architectures may
bring further improvement.  Finally, we could provide the neural networks with
more information about the premises. Currently, the premises are just names
(words) and NMT can only learn their \emph{latent
semantics}~\cite{DeerwesterDLFH90}.  Adding more information about their
logical representation and meaning may be useful.

\bibliographystyle{abbrv}
\bibliography{references-bartosz,ate11}

\appendix

\section{Subproofs as Standalone Data Set}
\label{sec:subproofs}

From all the proofs in the initial data set (not only the training part) we
extracted 60299 different pairs of the form
\[
	\big(
	\texttt{lemma},
	\{\texttt{premise\_1}, \texttt{premise\_2}, \ldots, \texttt{premise\_n}\}
	\big),
\]
same as those used for augmenting the training set (Section
\ref{sec:augmentation_with_subproofs}). This means that:
\begin{itemize}
	\item \texttt{lemma} is an intermediate sublemma appearing in the compacted
	representation of the proof (see Section \ref{sec:target_order} and Figure
	\ref{fig:tree}), which has no negated conjecture of the original proof
	among its ancestors,
	\item $\{\texttt{premise\_1}, \texttt{premise\_2}, \ldots, \texttt{premise\_n}\}$
		is a set of all the premises being among ancestors of the \texttt{lemma}.
\end{itemize}

These pairs have 29616 different lemmas (each lemma may have several
different proofs). We split these lemmas into training and testing parts in
proportions approximately 0.75 and 0.25, respectively (independently from the main
training/testing split of the Mizar theorems). We also recorded information
about the heights of the subproof trees from which 
the lemmas were
extracted -- to investigate how the height of the tree correlates with the
difficulty of learning premise selection. The heights of the subtrees vary
between 1 and 35, where lower subtrees are much more frequent than higher ones.

The NMT system was trained on the training examples, with the same
settings as in the main experiments. Additionally, we trained the XGBoost
system for comparison.

The results of the evaluation on the testing examples, in terms of the similarity
metrics (Jaccard index and Coverage) as well as an ATP evaluation, are presented
in Table \ref{tab:results_subproofs}. The table presents the performance of the
machine learning methods with respect to all the testing examples as well as on
subsets of them selected according to the heights of subtrees of proof trees
a given sublemma originated from.

Overall, in comparison to the main data, premise selection for subproofs
turned out to be significantly easier task for both machine learning methods.
On the whole testing set the ATP performance was 83\% for NMT and 61\% for
XGBoost. When running the automated prover without any premise selection
advice, with all available premises as axioms, the ATP success rate was
13\%.

As for the results depending on the heights of the subtrees, in the table we
present them up to the height 9 -- for larger values the number of lemmas
becomes very small. There are two trends visible for both NMT and XGBoost: with
increased height the Jaccard metric goes up and Coverage goes down. The likely
explanation is that the lower trees contain less premises in their leaves and
precise selection of them by the predictor is less likely, hence the low Jaccard
metric. On the other hand, the higher trees have more premises in their
leaves and covering them by the predictor is more difficult, which is reflected by the
low Coverage. The dependence of ATP performance on the heights is unclear.
Surprisingly, it is \emph{not} the case that the smallest subtrees contained the
easiest premise selection examples. 

\begin{table}
\caption{\label{tab:results_subproofs}
Performance of the NMT and XGBoost models evaluated on the testing part of the
subproofs data set. We use our similarity metrics (Jaccard index and Coverage)
and ATP evaluation (columns named Proved). The first column contains information
about the average height of the proof subtrees the sublemmas originated from.
The second column is the number of lemmas in a given subset.
The first numeric row refers to all the testing examples, independently of
the height. The largest numbers in the columns are marked in bold.
}
\centering
\begin{tabular}{cc@{\hspace{0.3cm}}ccc@{\hspace{0.3cm}}ccc@{\hspace{0.1cm}}}
\toprule
	&&\multicolumn{6}{c}{\textbf{Machine learning model}} \\
\cmidrule[0.05em](r){3-8}
&&\multicolumn{3}{c}{\texttt{NMT}} & \multicolumn{3}{c}{\texttt{XGBoost}} \\
	\cmidrule[0.05em](r){3-5} \cmidrule[0.05em](r){6-8}
	\textbf{Height} & \#Lemmas & Jaccard & Coverage & Proved & Jaccard & Coverage & Proved \\
\midrule
$[1,\infty)$ & 7300 & 0.30 & 0.80 & 0.83  & 0.27  & 0.74  & 0.61  \\
\midrule
	$1$  & 1610 & 0.19 & 0.84 & 0.83 & 0.18 & \textbf{0.82} & 0.66 \\
	$(1,2]$  & \textbf{1803} & 0.29 & \textbf{0.87} & 0.84 & 0.25 & 0.79 & 0.60 \\
$(2,3]$  & 1431 & 0.34 & 0.82 & 0.81 & 0.30 & 0.75 & 0.57 \\
$(3,4]$  & 936  & 0.36 & 0.75 & 0.82 & 0.31 & 0.69 & 0.56 \\
$(4,5]$  & 580  & 0.34 & 0.70 & 0.83 & 0.31 & 0.64 & 0.60 \\
$(5,6]$  & 319  & 0.37 & 0.70 & 0.86 & 0.33 & 0.64 & 0.62 \\
$(6,7]$  & 223  & 0.34 & 0.64 & 0.83 & 0.32 & 0.60 & 0.64 \\
	$(7,8]$  & 124  & 0.37 & 0.70 & 0.89 & 0.34 & 0.66 & \textbf{0.79} \\
	$(8,9]$  & 90   & \textbf{0.40} & 0.69 & \textbf{0.92} & \textbf{0.36} & 0.63 & 0.71 \\
\bottomrule
\end{tabular}
\end{table}

\section{Examples of predictions from RNN}
\label{sec:examples}

In this section we show two examples of predictions from the
recurrent neural NMT system and compare them with the respective
XGBoost predictions. All the presented predictions come from the
systems trained on the \texttt{basic} data set. Note that in both
cases below, the NMT predictions are more diverse, expressing
different proof approaches and allowing quite different proof
attempts.  On the other hand, as soon as XGBoost ranks high a bad set
of lemmas that mislead E prover, adding more premises does not help in
these cases.

\subsection{Theorem \texttt{t128\_zfmisc\_1}}

\vspace{2mm}
\para{Mizar statement of the theorem:}
\vspace{-1mm}
\begin{verbatim}
for x, y, z, Y being set holds
( [x,y] in [:{z},Y:] iff ( x = z & y in Y ) )
\end{verbatim}

\vspace{2mm}
\para{NMT predictions:}
\vspace{-1mm}
\begin{footnotesize}
\begin{verbatim}
1: d1_enumset1 t71_enumset1 t69_enumset1 t70_enumset1 l54_zfmisc_1
2: d2_tarski t77_enumset1 t79_enumset1 t76_enumset1 t84_enumset1 l54_zfmisc_1
3: l38_zfmisc_1 t69_enumset1 t70_enumset1 t71_enumset1 t20_zfmisc_1 l54_zfmisc_1
4: d1_tarski t69_enumset1 l54_zfmisc_1 d2_tarski
5: d1_tarski t69_enumset1 l54_zfmisc_1
6: l33_zfmisc_1 t69_enumset1 t70_enumset1 t71_enumset1 t20_zfmisc_1 l54_zfmisc_1
7: t76_enumset1 d1_enumset1 l54_zfmisc_1
8: t20_zfmisc_1 t69_enumset1 t70_enumset1 t71_enumset1 t65_zfmisc_1 l54_zfmisc_1
9: d2_tarski t70_enumset1 t71_enumset1 t69_enumset1 l54_zfmisc_1
10: l24_zfmisc_1 t69_enumset1 t70_enumset1 t71_enumset1 d1_enumset1 l54_zfmisc_1
\end{verbatim}
\end{footnotesize}

\vspace{2mm}
\para{XGBoost predictions (ranking):}
\vspace{-1mm}
\begin{footnotesize}
\begin{verbatim}
Ranking: d1_tarski t69_enumset1 t70_enumset1 t71_enumset1 l54_zfmisc_1 t106_zfmisc_1 t77_enumset1 d3_tarski ...
\end{verbatim}
\end{footnotesize}

\vspace{2mm}
\para{Comparison:}
E prover without \texttt{auto} mode was able to prove \texttt{t128\_zfmisc\_1}
with 5th predictions proposed by NMT:
\begin{verbatim}
5: d1_tarski t69_enumset1 l54_zfmisc_1
\end{verbatim}
but no proof could be found with top slices
of the ranking proposed by XGBoost. The reason for this is that premises appearing in the top part
of the ranking:
\begin{verbatim}
t69_enumset1 t70_enumset1 t71_enumset1
\end{verbatim}
are very similar, and E prover is stuck with them, even with higher CPU time limits.
Below there are the Mizar statements of the discussed premises:
\vspace{-2mm}
\begin{verbatim}
d1_tarski: for x being set holds ( x in it iff x = y );
l54_zfmisc_1: [x,y] in [:X,Y:] iff x in X & y in Y
t69_enumset1: for x1 being set holds {x1,x1} = {x1}
t70_enumset1: for x1, x2 being set holds {x1,x1,x2} = {x1,x2}
t71_enumset1: for x1, x2, x3 being set holds {x1,x1,x2,x3} = {x1,x2,x3}
\end{verbatim}

\subsection{Theorem \texttt{t30\_tops\_1}}

\vspace{2mm}
\para{Mizar statement of the theorem:}
\vspace{-1mm}
\begin{verbatim}
for GX being TopStruct
for R being Subset of GX holds
( R is open iff R ` is closed )
\end{verbatim}

\vspace{2mm}
\para{NMT predictions:}
\vspace{-1mm}
\begin{footnotesize}
\begin{verbatim}
1: dt_k3_subset_1 d1_tops_1 t52_pre_topc t29_tops_1 d8_tops_1 d7_tops_1
2: t100_xboole_1 t12_setfam_1 t28_xboole_1 t48_xboole_1 commutativity_k2_tarski d5_subset_1 ...
3: involutiveness_k3_subset_1 t29_tops_1 t101_tops_1 t52_pre_topc dt_k3_subset_1 d8_tops_1 d7_tops_1
4: t100_xboole_1 t12_setfam_1 t36_xboole_1 t48_xboole_1 t7_ordinal1 t2_xboole_1 d5_xboole_0 ...
5: t100_xboole_1 t12_setfam_1 d5_subset_1 t2_boole t91_tops_1 involutiveness_k3_subset_1 dt_k3_subset_1 ...
6: t28_xboole_1 t12_setfam_1 commutativity_k3_xboole_0 t100_xboole_1 t22_xboole_1 t36_xboole_1 ...
7: t12_setfam_1 t70_enumset1 t100_xboole_1 d5_subset_1 t71_enumset1 t72_enumset1 t73_enumset1 t74_enumset1 ...
8: d10_xboole_0 t2_xboole_1 t43_xboole_1 t12_xboole_1 commutativity_k2_xboole_0 t41_xboole_1 t36_xboole_1 ...
9: d10_xboole_0 t2_xboole_1 t43_xboole_1 t12_xboole_1 commutativity_k2_xboole_0 t41_xboole_1 t36_xboole_1 ...
10: d10_xboole_0 t2_xboole_1 t43_xboole_1 t12_xboole_1 commutativity_k2_xboole_0 t41_xboole_1 t36_xboole_1 ...
\end{verbatim}
\end{footnotesize}

\vspace{2mm}
\para{XGBoost predictions (ranking):}
\vspace{-1mm}
\begin{footnotesize}
\begin{verbatim}
Ranking: d10_xboole_0 t3_subset d4_subset_1 d3_struct_0 d5_subset_1 involutiveness_k3_subset_1 dt_k3_subset_1
     dt_k2_pre_topc dt_l1_pre_topc d3_tarski dt_k2_subset_1 redefinition_k7_subset_1 ...
\end{verbatim}
\end{footnotesize}

\vspace{2mm}
\para{Comparison:}
E prover without \texttt{auto} mode was able to prove \texttt{t30\_tops\_1}
with 3rd predictions proposed by NMT:
\begin{footnotesize}
\begin{verbatim}
3: involutiveness_k3_subset_1 t29_tops_1 t101_tops_1 t52_pre_topc dt_k3_subset_1 d8_tops_1 d7_tops_1
\end{verbatim}
\end{footnotesize}
actually using these 3 premises:
\begin{verbatim}
t29_tops_1 involutiveness_k3_subset_1 dt_k3_subset_1
\end{verbatim}
E prover was not able to prove the theorem with any top slice of the ranking.

\end{document}